%% file: report.tex
\documentclass[11pt]{elsarticle}
\usepackage[margin=1in]{geometry}
\usepackage{mathtools}
\usepackage{fullpage}

\input{inputs}

\title{A C implementation of the Smith massager algorithm}

\author{Ziwen Wang}
\address{Cheriton School of Computer Science, University of Waterloo,
Waterloo ON, Canada N2L 3G1}
\ead{zi.wen@uwaterloo.ca}

\author{Stavros Birmpilis}
\address{Cheriton School of Computer Science, University of Waterloo,
Waterloo ON, Canada N2L 3G1}
\ead{sbirmpilis@uwaterloo.ca}

\author{George Labahn}
\address{Cheriton School of Computer Science, University of Waterloo,
Waterloo ON, Canada N2L 3G1}
\ead{glabahn@uwaterloo.ca}

\author{Arne Storjohann}
\address{Cheriton School of Computer Science, University of Waterloo,
Waterloo ON, Canada N2L 3G1}
\ead{astorjoh@uwaterloo.ca}
\date{}

\begin{document}
\maketitle

\input{abstract}

\input{introduction}

\input{algorithm}

\input{implementation}

\input{experiments}

\input{conclusion}

\bibliographystyle{plain}
\newcommand{\SortNoop}[1]{}

\end{document}

%% file: inputs.tex
\usepackage{latexsym}
\usepackage{amssymb}
\usepackage{amsmath}
\usepackage{amsthm}
\usepackage{color}
\usepackage{enumitem}
\usepackage{hyperref}
\usepackage{booktabs}
\usepackage{graphicx}
\usepackage{pgfplots}
\pgfplotsset{compat=1.16}

\newcommand{\M}{{\mathsf{M}}}
\newcommand{\B}{{\mathsf{B}}}

\newcommand{\sO}[2]{{#1}^{#2+o(1)}}

\newcommand{\Z}{\ensuremath{\mathbb Z}}

\newcommand{\diag}{{\rm diag}}
\newcommand{\Znn}{\ensuremath{\mathbb Z}^{n\times n}}

\newcommand{\mylabel}[1]{\label{#1}}

\newtheorem{theorem}{Theorem}
\newtheorem{definition}[theorem]{Definition}

\newtheorem{remark}[theorem]{Remark}

\newlength{\algwidth}


%% file: abstract.tex
\begin{abstract}
We describe a C implementation of the Las Vegas algorithm
of~\cite{BirmpilisLabahnStorjohann20} for computing the Smith normal
form of a nonsingular integer matrix.  The algorithm computes a Smith
massager for the input matrix using $O(n^{\omega}\, \B(\log n + \log
\|A\|)\, (\log n)^2)$ bit operations, which is softly equivalent to
the cost of multiplying two matrices of the same dimension and entry
size.  We describe the key implementation techniques that bridge the
gap between the theoretical algorithm and practical performance,
including BLAS-accelerated modular arithmetic via the Residue Number
System and an adaptive batching scheme that collapses the theoretical
$O(\log n)$ iterations to $O(1)$ in practice.  Experiments on
matrices of dimension up to $n = 10007$ show that the implementation's
running time scales proportionally to that of a single BLAS matrix
multiplication, with both exhibiting the same effective growth rate on
a log-log plot.
\end{abstract}

%% file: introduction.tex
\section{Introduction}

Let $A \in \Znn$ be a nonsingular integer matrix. Then there exists a
unique diagonal matrix $S = \diag(s_1, s_2, \ldots, s_n)$  with 
$s_1 \mid s_2 \mid \cdots \mid s_n$ and  unimodular matrices $U,V \in \Znn$
such that $UAV = S$. The matrix $S$ is the Smith normal form for $A$ while $U$ and $V$ represent
row and column operations invertible over the integers converting $A$ to $S$. 
The Smith normal form is a fundamental object in
integer linear algebra with applications to finite abelian group
classification~\cite{Cohen,Newman}, linear system
solving~\cite{BirmpilisLabahnStorjohann21}, lattice computations, and
many other areas.

A natural goal is to design algorithms for the Smith form problem with
cost softly equivalent to that of multiplying two matrices of the same
dimension and entry size as the input.  If $\omega$ denotes a valid
exponent for matrix multiplication --- two $n \times n$ matrices can
be multiplied in $O(n^\omega)$ operations from the domain of entries
--- and $\|A\| = \max_{ij} |A_{ij}|$ is the largest entry in absolute
value, then the target complexity is $\sO{(n^\omega \log \|A\|)}{1}$
bit operations.  Following~\cite{vonzurGathenGerhard}, we use $\M(d)$
to bound the cost of multiplying two integers bounded by $2^d$, and
$\B(d) = O(\M(d) \log d)$ for integer gcd-related computations.
In~\cite{BirmpilisLabahnStorjohann19}, integer linear system solving
was deterministically reduced to matrix multiplication.
Building on this,~\cite{BirmpilisLabahnStorjohann20} gave a Las Vegas
algorithm for the Smith form that achieves the target complexity up to
logarithmic factors:
$O(n^\omega\, \B(\log n + \log \|A\|)\, (\log n)^2)$ bit operations.
The central tool introduced in that paper is the \emph{Smith
massager}, a compact representation of $A^{-1}$ modulo the invariant
factors.  A subsequent paper~\cite{BirmpilisLabahnStorjohann21}
extended the approach to compute the unimodular multipliers as well.

\paragraph{Our contributions.}
In this paper we describe a C implementation of the Smith massager
algorithm.  The implementation is built on top of the GMP
library~\cite{GMP} for
arbitrary-precision arithmetic, the FLINT library~\cite{FLINT} for
integer matrix operations, and OpenBLAS~\cite{OpenBLAS} for fast
numerical linear algebra.  The IML library~\cite{IML}
provides certified rational system solving.  We detail the key
implementation techniques that bridge the gap between the theoretical
algorithm and practical performance 
and
present experimental results 
demonstrating
that the observed running time grows proportionally to matrix
multiplication cost, as predicted by the theory.

\paragraph{Organization.}
This paper is organized as follows. In 
Section~\ref{sec:alg} we give a brief  overview of the Smith massager algorithm from~\cite{BirmpilisLabahnStorjohann20}.  Our implementation and design choices are  then described in Section~\ref{sec:impl} .  In  Section~\ref{sec:exp} we present our  experimental results including
correctness verification and asymptotic scaling analysis.
The paper  ends with a conclusion and topics for future research.

%% file: algorithm.tex
\section{Algorithm overview}
\mylabel{sec:alg}

In this section we give a self-contained overview of the Smith massager algorithm
from~\cite{BirmpilisLabahnStorjohann20}.  For complete proofs, we
refer the reader to that paper and
to~\cite{BirmpilisLabahnStorjohann21}.

\subsection{The Smith massager}

The Smith massager is best defined as follows.
\begin{definition}[\protect{\cite[Definition~1]{BirmpilisLabahnStorjohann21}}]
 \mylabel{def:sm}
Let $A\in\Znn$ be nonsingular with Smith form $S$.
The pair of matrices $(S, F)$ is a  \emph{Smith massager} for $A$ if
$F \in \Znn$ and
\begin{itemize}
\item[(i)]
$AF \equiv 0 \bmod S$, and
\item[(ii)] there exists a matrix $W \in\Znn$ such that
$W F \equiv I_n \bmod S.$
\end{itemize}
\end{definition}

\begin{remark}
Note that condition (ii) in Definition~\ref{def:sm} is equivalent to
$S$ and $F$ being right coprime.
\end{remark}

In~\cite{BirmpilisLabahnStorjohann20} the Smith massager was defined as
a tuple $(U, M, T, S)$ of $n \times n$ integer matrices such that the
$2n \times 2n$ matrix
\begin{equation} \mylabel{eq:Bdef}
\setlength{\arraycolsep}{.8\arraycolsep}
\renewcommand{\arraystretch}{.9}
B := \left [ \begin{array}{c|c} A & \\\hline
 & I_n \end{array} \right ]
\left [ \begin{array}{c|c} I_n & \\\hline
U & I_n \end{array} \right ]
\left [ \begin{array}{c|c} I_n & M \\\hline
  & T \end{array} \right ]
\left [ \begin{array}{c|c} I_n &   \\\hline
  & S^{-1} \end{array} \right ]
\end{equation}
is integral. Here $T$ is unit upper triangular and $S$ is nonsingular
and in Smith form.
In this form the massager was said to be \emph{maximal} if $S$ is the Smith form of~$A$.
Since~(\ref{eq:Bdef}) implies $(\det B)(\det S) = \det A$, the
massager is maximal if and only if $B$ is unimodular.  The total
storage for a reduced Smith massager $(U, M, T, S)$ is $O(n^2 (\log n +
\|A\|))$ bits, essentially the same as required to represent the input matrix.

The two definitions are equivalent by observing that $W = T^{-1}  U$. Definition \ref{def:sm} captures the basic concepts of a massager
while the original definition was useful for describing the Las Vegas algorithm in~\cite{BirmpilisLabahnStorjohann20}.  There the notion of
an index-$(m,r)$ Smith massager plays a central role.

\begin{definition}[Index-$(m,r)$ Smith massager]
\label{def:ism}
Let $B \in \Z^{2n \times 2n}$ be nonsingular with the shape
$$
B = \left [ \begin{array}{ccc} A & & \ast \\
 & I_{n-m} & \\
\ast & & \ast \end{array} \right ].$$
For $m, r \in \Z_{\geq  0}$ such that $m+r \leq n$, an \emph{index-$(m,r)$
Smith massager for $B$} is a tuple $(U, M, T, S) \in
(\Z^{r \times n}, \Z^{n \times r}, \Z^{r \times r}, \Z^{r \times
r})$ such that the matrix
\begin{equation} \label{eq:barB1}
\setlength{\arraycolsep}{.6\arraycolsep}
\renewcommand{\arraystretch}{.8}
C := B
\left [ \begin{array}{cccc} I_n & & & \\
 & I & & \\
U & & I_r & \\
 & & & I_m \end{array} \right ]
\left [ \begin{array}{cccc} I_n & & M & \\
 & I & & \\
  & & T & \\
 & & & I_m \end{array} \right ]
\left [ \begin{array}{cccc} I_n & &  & \\
 & I & & \\
  & & S^{-1} & \\
 & & & I_m \end{array} \right ]
\end{equation}
is integral, with $S$ nonsingular and in Smith form, and $T$ unit
upper triangular. We say that $(U,M,T,S)$ is \emph{maximal} for
$B$ if $S$ is comprised of the $r$ largest invariant factors of the
Smith form of $B$.
\end{definition}
\noindent
When $m=0$ the matrix $B$ is equal to $\diag(A,I_n)$.
If in addition, $r=n$, then an index-$(m,r)$ Smith massager for
$\diag(A,I_n)$ corresponds to a Smith massager for $A$ as defined
earlier.

\subsection{Three-phase algorithm}

The algorithm has three phases:

\paragraph{Phase 1: Largest invariant factor}
Compute the largest invariant factor $s_n$ of $A$ (or a positive
multiple thereof) using a Monte Carlo method.  This provides the
initial modulus $s$ for the iterative construction.

\paragraph{Phase 2: Iterative massager construction}
Starting from the trivial massager $(0, 0, I_n, I_n)$, the algorithm
performs $O(\log n)$ iterations.  At iteration $i$, the algorithm
extracts the next batch of $r_i$ invariant factors by computing an
\emph{index-$(m, r_i)$ Smith massager} for the current massaged matrix
$B$.  Each iteration involves:
\begin{enumerate}[itemsep=2pt]
\item Computing a random projection $P = \text{Rem}(s B^{-1} J,\, s)$
  for a randomly chosen $J \in \Z^{2n \times (r_i + k)}$, via an
  integrality certification procedure.  This reduces the problem to
  working modulo the current $s$.
\item Computing the modular Smith form of the $n \times r_i$ projection
  $P_1$ over $\Z/(s)$, yielding matrices $U \in \Z^{r_i \times n}$
  and $M \in \Z^{n \times r_i}$ such that $-UP_1 \equiv D \pmod{s}$
  and $P_1 \equiv MD \pmod{s}$, where $D$ is the reverse Smith form
  of $P_1$ modulo $s$.
\item Setting $S' = sD^{-1}$ (the extracted invariant factors) and
  updating the massaged matrix $B$ accordingly.
\end{enumerate}
The batch sizes $r_i$ grow geometrically: $r_0, 2r_0, 4r_0, \ldots$,
exploiting the ``dimension $\times$ precision $\leq$ invariant''
compromise~\cite{Storjohann04}.  As more invariant factors are
extracted, the working modulus $s$ shrinks, so the cost per
extracted factor decreases.

\paragraph{Phase 3: Unimodularity certification}
Verify that the final massaged matrix $B$ is unimodular using the
deterministic algorithm of~\cite{PauderisStorjohann12}.  The
randomized projection in Phase~2 may fail to produce a maximal
massager (with probability at most $1/2$); if Phase~3 detects that
$B$ is not unimodular, the entire algorithm restarts from Phase~1.

\subsection{Cost analysis}

The cost of Phase~2 is dominated by $O(\log n)$ calls to the index
massager, each involving an integrality certification and a modular
Smith form computation.  The integrality certification for an $n
\times n$ matrix with modulus $s$ costs $O(n^\omega\, \M(\log s)\,
\log n)$ bit operations.  Summing over all iterations, and using the
dimension--precision tradeoff, the total cost is $O(n^\omega\, \B(\log
n + \log \|A\|)\, (\log n)^2)$ bit operations.

\begin{theorem}[\cite{BirmpilisLabahnStorjohann20}]
There exists a Las Vegas algorithm that computes the Smith form of a
nonsingular $A \in \Znn$ using $O(n^\omega\, \B(\log n + \log \|A\|)\,
(\log n)^2)$ bit operations.
\end{theorem}

%% file: implementation.tex
\section{Implementation techniques}
\mylabel{sec:impl}

The implementation is written in C and depends on
GMP~\cite{GMP} for arbitrary-precision arithmetic,
FLINT~\cite{FLINT} for integer matrix operations, and
OpenBLAS~\cite{OpenBLAS} for BLAS-accelerated numerical linear
algebra.  The IML library~\cite{IML} provides certified rational
system solving via Dixon's $p$-adic lifting.

The main algorithm and the RNS arithmetic infrastructure live under
\texttt{src/core/}.  Two supporting modules,
\texttt{src/lift/} (double-plus-one $p$-adic
lifting~\cite{Storjohann04}) and \texttt{src/spinv/}
(unimodularity certification via high-order
residues~\cite{PauderisStorjohann12}), implement standard
techniques from the literature; since their implementation follows the
published descriptions, we do not discuss their internals further.

In this section we document the engineering choices made in our
implementation, which are important for practical performance.

\subsection{BLAS-accelerated modular arithmetic via the Residue Number System}
\mylabel{ssec:rns}

The dominant cost in the algorithm is integer matrix multiplication.
The Residue Number System (RNS) reduces multi-precision multiplication
of $n \times n$ matrices with $d$-bit entries from $O(n^3 \, \M(d))$
to $O(n^3 \cdot d)$ bit operations by mapping the work to word-sized
BLAS calls -- the standard approach used by IML~\cite{IML} and other
libraries for exact integer linear algebra.

An integer matrix $A$ with entries of bitlength $d$ is reduced modulo
$\ell$ coprime word-sized primes, producing $\ell$ \emph{residue
matrices}.
The product $A \cdot B$ is computed as $\ell$ independent
double-precision matrix multiplications using OpenBLAS's
\texttt{cblas\_dgemm}, and the integer result is reconstructed via the
Chinese Remainder Theorem.

Our specific choices:
\begin{itemize}[itemsep=2pt]
\item \textbf{Exact double-precision accumulation.}  For the BLAS
  output to be exact, the accumulated inner product must not exceed
  $2^{53} - 1$.  The largest safe modulus $p$ for dimension $n$ is
  determined from $n(p-1)^2 + (p-1) < 2^{53} - 1$.  Primes are
  chosen consecutively below this bound.

\item \textbf{Fast modular reduction.}  The inner-loop reduction
  $r = a - \lfloor a / p \rfloor \cdot p$ avoids the expensive
  integer division instruction by using multiplication by a
  precomputed floating-point reciprocal $1/p$.

\item \textbf{Fast RNS base conversion.}  Converting a matrix from
  one RNS basis to another (needed during lifting) is done with a
  Lagrange-based shortcut that uses a floating-point correction
  factor, bypassing the expensive CRT reconstruction step entirely.

\item \textbf{Precomputed RNS constants.}  The RNS basis stores
  precomputed Lagrange interpolants, modular inverses, and the
  floating-point weights used by the base conversion above.  These
  are computed once and reused across all matrix operations.
\end{itemize}

\subsection{Adaptive batch sizing}
\mylabel{ssec:adaptive}

The theoretical algorithm extracts the invariant factors in
$O(\log n)$ batches of geometrically growing size ($r_0, 2r_0, 4r_0,
\ldots$).  The implementation collapses this to only 3 or 4
iterations -- a strategy also adopted in~\cite{PauderisStorjohann13},
there with a fixed coarse schedule.  Here, the size of each batch is
chosen adaptively based on how fast the working modulus $s$ shrank in
the previous iteration, and once $s$ drops below word size the final
iteration absorbs all remaining invariant factors at once.  Each batch performs an integrality certification via $p$-adic
lifting, whose initialization cost (computing $A^{-1} \bmod p$) is
comparable to the lifting work itself.  Collapsing many small
batches into 3 or 4 large ones amortizes this fixed cost over much
more work -- a significant speedup in practice.

%% file: experiments.tex
\section{Experimental results}
\mylabel{sec:exp}

All experiments were conducted on an Apple M4 Max processor (ARM64)
with 64~GB RAM, running macOS~26.4.  The implementation was compiled
with Apple Clang using OpenBLAS (Homebrew), GMP~6, and FLINT~3.

\subsection{Test matrices}

For each prime $n$, we construct the $n \times n$ matrix $A$ with
entries $A_{ij} = i^j \bmod n$ for $0 \leq i, j < n$.  This is a
Vandermonde matrix with entries reduced modulo $n$, so $\log \|A\| =
O(\log n)$.  Primality of $n$ ensures $A$ is nonsingular.  Following
J\"ager and Wagner~\cite{JagerWagner}, we use this family of matrices because its
Smith form is highly nontrivial: $A$ typically has more than $n/2$
nontrivial invariant factors, and the largest is very large relative
to $n$.  For instance, at $n=1009$ the largest invariant factor has
4083 bits.  For these matrices the theoretical cost reduces to $O(n^\omega
\, \B(\log n) \, (\log n)^2)$, and since OpenBLAS uses standard cubic
matrix multiplication ($\omega = 3$), this is $O(n^3 \, (\log n)^c)$
for a small constant $c$.

\subsection{Correctness verification}

For each test prime $n \leq 53$, we verify the returned Smith
massager $(U, M, T, S)$ against four independent checks:
\begin{enumerate}[itemsep=2pt]
\item \textbf{Divisibility:} $s_1 \mid s_2 \mid \cdots \mid s_n$.
\item \textbf{Smith form:} The invariant factors of $S$ agree with an
  independent computation using SymPy~\cite{SymPy}.
\item \textbf{Integrality:} The massaged matrix $B$
  in~(\ref{eq:Bdef}) is integral, verified by checking that
  $AMS^{-1}$ and $(T + UM)S^{-1}$ have integer entries.
\item \textbf{Unimodularity:} $|\det B| = 1$, verified by checking
  $\det A / \prod_i s_i = \pm 1$.
\end{enumerate}
All primes from 3 to 53 pass all four checks.
For larger dimensions ($n$ up to 10007), Phase~3 of the
algorithm certifies unimodularity; all test cases succeeded.  The verification
script is included in the repository at
\texttt{tests/verify\_sympy.py}.

\subsection{Running time}

In all trials reported here the algorithm succeeded on the first
attempt (no Las~Vegas restarts).  Table~\ref{tab:timing} reports
the median wall-clock time over 3 trials for each dimension.

\begin{table}[ht]
\centering
\caption{Wall-clock times (seconds) for the Smith massager algorithm,
median over 3~trials.}
\mylabel{tab:timing}
\smallskip
\begin{tabular}{rrr}
\toprule
$n$ & Time (s) & $10^9 \cdot t/n^3$ \\
\midrule
53    & 0.033   & 222 \\
101   & 0.083   & 81 \\
211   & 0.447   & 47 \\
503   & 2.07    & 16 \\
1009  & 26.30   & 26 \\
2003  & 134.4   & 17 \\
5003  & 1744    & 14 \\
10007 & 5399    & 5.4 \\
\bottomrule
\end{tabular}
\end{table}

The last column shows the ratio $t/n^3$, which decreases across the
test range.  This is a pre-asymptotic effect: BLAS \texttt{dgemm} (matrix multiplication)
itself becomes more efficient as $n$ grows, so its cost per $n^3$
operation also drops in this regime.

\subsection{Scaling analysis}

To assess asymptotic behavior, we compare the Smith massager's
running time against the cost of a single double-precision
$n \times n$ matrix multiplication using an optimized BLAS
(\texttt{cblas\_dgemm}).  If the Smith massager cost scales as
$n^\omega$ times slowly-growing factors, both curves should be
parallel on a log-log plot.

Figure~\ref{fig:loglog} shows this comparison.  Both curves grow at
the same rate, with an approximately constant vertical gap.  This confirms that the
Smith massager cost is proportional to matrix multiplication cost, as
predicted by the theory.

\begin{figure}[ht]
\centering
\begin{tikzpicture}
\begin{loglogaxis}[
    xlabel={Dimension $n$},
    ylabel={Time (seconds)},
    width=0.85\textwidth,
    height=0.55\textwidth,
    grid=major,
    legend pos=north west,
    legend style={font=\small},
    xmin=40, xmax=15000,
    ymin=1e-6, ymax=10000,
]
\addplot[only marks, mark=*, mark size=2pt, blue] coordinates {
    (53, 0.033) (101, 0.083) (211, 0.447)
    (503, 2.07) (1009, 26.30) (2003, 134.4)
    (5003, 1744) (10007, 5399)
};
\addlegendentry{Smith massager}

\addplot[only marks, mark=triangle*, mark size=2pt, red] coordinates {
    (53, 0.000009) (101, 0.000094) (211, 0.000186)
    (503, 0.001219) (1009, 0.005711) (2003, 0.039892)
    (5003, 0.433997) (10007, 3.378873)
};
\addlegendentry{\texttt{dgemm} ($n \times n$ doubles)}

\addplot[domain=40:15000, samples=50, black!50, dotted, thick] {exp(-16.5)*x^3};
\addlegendentry{$n^{3}$ (reference)}

\end{loglogaxis}
\end{tikzpicture}
\caption{Log-log plot of Smith massager time and BLAS \texttt{dgemm}
  time versus dimension.  Both curves track each other with a vertical
  gap of roughly $10^3$ across the range.  The dotted line shows
  $n^3$ for reference.}
\mylabel{fig:loglog}
\end{figure}
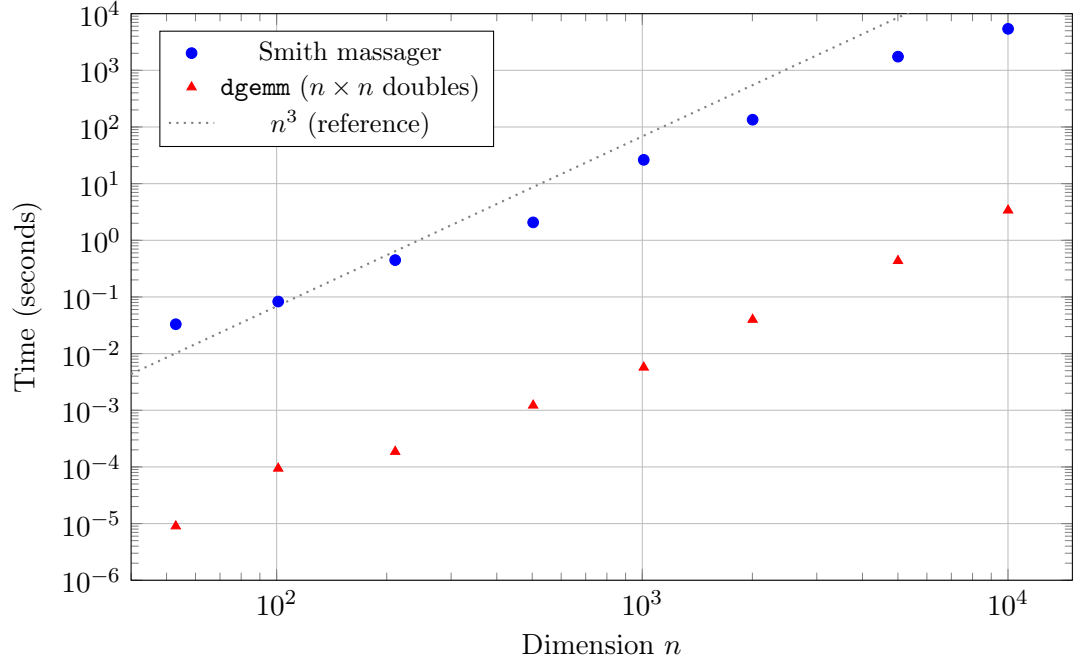

The vertical gap reflects the constant-factor overhead of the Smith
massager over a single \texttt{dgemm} call: the algorithm performs
$O(\ell \cdot (\log n)^c)$ BLAS calls per run, where $\ell \approx
3$--$4$ is the number of RNS primes.

\subsection{Internal timing breakdown}

Table~\ref{tab:breakdown} shows the internal timing breakdown for
selected dimensions.

\begin{table}[ht]
\centering
\caption{Internal timing breakdown (seconds) from a single run,
reported by the implementation's built-in timers.
LIF = largest invariant factor (Phase 1),
IM = index massager iterations (Phase 2),
UC = unimodularity certification (Phase 3).}
\mylabel{tab:breakdown}
\smallskip
\begin{tabular}{rrrrr}
\toprule
$n$ & LIF & IM & UC & Total \\
\midrule
53   & 0.011   & 0.005   & 0.004   & 0.020 \\
101  & 0.024   & 0.026   & 0.018   & 0.070 \\
211  & 0.078   & 0.263   & 0.078   & 0.428 \\
503  & 0.644   & 0.762   & 0.563   & 2.069 \\
1009 & 2.424   & 17.68   & 3.835   & 24.67 \\
2003 & 18.76   & 93.01   & 23.43   & 139.7 \\
5003 & 168.6   & 692.2   & 545.6   & 1515 \\
\bottomrule
\end{tabular}
\end{table}

%% file: conclusion.tex
\section{Conclusion}
\mylabel{sec:conc}

We have described a C implementation of the Las Vegas algorithm
of~\cite{BirmpilisLabahnStorjohann20} for computing the Smith normal
form via the Smith massager.  The implementation makes integer matrix
multiplication practical at scale through BLAS-accelerated RNS
arithmetic, and uses adaptive batch sizing to amortize the substantial
fixed cost of each batch's integrality certification.

Our experiments on matrices of dimension up to $n = 10007$ confirm the
theoretical prediction: the Smith massager's running time scales
proportionally to matrix multiplication cost, with both exhibiting
the same effective growth exponent on a log-log plot.  Correctness
is verified independently against SymPy for small inputs, together with the
unimodularity of the massaged matrix~$B$ certified for each run.

The source code is available at
\url{https://github.com/SmithMassager/SmithMassagerC}.